\documentclass[preprint,
superscriptaddress,
amsmath,amssymb,aps,showkeys,showpacs,
twoside,final,secnumarabic,
nofootinbib]{revtex4-2}

\usepackage[paperwidth=205mm,paperheight=290mm,top=17mm,bottom=25mm,
inner=17mm,outer=17mm,
twoside]{geometry}

\usepackage{cmap} 
\usepackage[T1,T2A]{fontenc}
\usepackage[utf8]{inputenc}
\usepackage[russian,english]{babel}

\makeatletter\AtBeginDocument{\let\@elt\relax}\makeatother

\usepackage{color}
\usepackage{graphicx}
\usepackage{dcolumn}
\usepackage{bm} 
\usepackage[unicode=true,colorlinks=true,linkcolor=magenta, urlcolor=blue, citecolor = blue,breaklinks]{hyperref}
\usepackage{multirow}
\usepackage{url}
\usepackage{breakurl}

\usepackage{comment}

\DeclareGraphicsExtensions{.eps}

\newcount\issue
\newcount\Vol
\newcount\numb
\usepackage{fancyhdr} 
\def\Vol{\textbf{80}}
\def\numb{x}
\begin{document}

\title{
Capability of using the normalizing flows\\ for extraction rare gamma events in the TAIGA experiment} 

\def\addressa{Skobeltsyn Institute of Nuclear Physics of M.V. Lomonosov Moscow State University, Moscow, 119991 Russia}
\def\addressb{Research Institute of Applied Physics of Irkutsk State University, Irkutsk, 664003 Russia}
\def\addressc{Institute for informatics and automation problems of the national Academy of Science of the Republic of Armenia, Yerevan, 0014 Republic of Armenia}

\author{\firstname{A.P.}~\surname{Kryukov}}
\email[E-mail: ]{kryukov@theory.sinp.msu.ru }
\affiliation{\addressa}
\author{\firstname{A.Yu.}~\surname{Razumov}}
\affiliation{\addressa}
\author{\firstname{A.P.}~\surname{Demichev}}
\affiliation{\addressa}
\author{\firstname{J.J.}~\surname{Dubenskaya}}
\affiliation{\addressa}
\author{\firstname{E.O.}~\surname{Gres}}
\affiliation{\addressb}
\affiliation{\addressa}
\author{\firstname{S.P.}~\surname{Polyakov}}
\affiliation{\addressc}
\author{\firstname{E.B.}~\surname{Postnikov}}
\affiliation{\addressa}
\author{\firstname{P.A.}~\surname{Volchugov}}
\affiliation{\addressa}
\author{\firstname{D.P.}~\surname{Zhurov}}
\affiliation{\addressb}
\affiliation{\addressa}


\begin{abstract}
The objective of this work is to develop a method for detecting rare gamma quanta against the background of charged particles in the fluxes from sources in the Universe with the help of the deep learning and normalizing flows based method designed for anomaly detection. It is shown that the suggested method has a potential for the gamma detection. The method was tested on model data from the TAIGA-IACT experiment. The obtained quantitative performance indicators are still inferior to other approaches, and therefore possible ways to improve the implementation of the method are proposed. 

\end{abstract}

\par
\keywords{gamma rays,  cosmic rays, extensive air showers, deep learning methods, rare events, normalizing flows   \\[5pt]}

\maketitle
\thispagestyle{fancy}


\section{Introduction}\label{intro}

Studying gamma rays of various energies generated in the vicinity of galactic and metagalactic sources is one of the most promising ways to study the sources themselves, and hence the important processes occurring in the Universe. The fact is that charged cosmic rays (elementary particles and atomic nuclei) are significantly influenced by galactic and intergalactic magnetic fields, which leads to a strong distortion of their trajectories and, as a result, to the loss of any information about the place of their origin. Gamma-ray astronomy does not have these shortcomings, since photons trajectories are not distorted because of their electric neutrality and therefore indicate the direction of their origin. In many ways, these reasons have contributed in recent years to the rapid development of experimental gamma-ray astronomy in the world \cite{1, 2}.

The Imaging Atmospheric Cherenkov Telescopes (IACTs) capture images of extensive air shower (EAS) generated by gamma rays and cosmic rays (charged particles) as they interact with the atmosphere. These images make it possible to draw conclusions about the properties of the primary particles.

In the past years, along with the previously developed special methods for identifying events with gamma rays and reconstructing their characteristics, methods based on deep learning have been successfully used \cite{2,3,4,5,6}. An important fact that must be taken into account is that the flux of gamma rays is very small compared to the flux of cosmic rays, the ratio is not higher than $\sim\,1:1000$, and therefore the success of IACT depends to a large extent on the ability to distinguish between these two types of events. Therefore its is crucial that the methods for classification of the recorded events perform well. For this aim new algorithms are explored to enhance the separation of gamma rays and charged cosmic rays including those based on deep learning \cite{2}. In this work we examine one more method to achieve this goal by using normalizing flow (NF) based generative models \cite{7}.

Among other applications, NFs were successfully used for anomaly detection; see, e.g., \cite{8} and refs. therein. However, gamma events are not anomalous but rather rare events. From the point of view of the machine learning, the crucial distinctions between anomalous and rare events are in the sizes of classes for training and testing sets. In the case of anomalies, as matter of rule, one does not know anything about them or at most quite a little. So that in the training set the anomaly class is empty or extremely small in comparison with the class of normal events. At the same time in the testing set the both class sizes can be approximately equal. On the contrary, in the case of rare events the training set contains classes of more or less equal sizes while in the testing one the class of rare events is extremely small. Therefore, one-class learning is generally used in case of anomalies, while two-class learning is generally used in case of rare events. In the field of the astroparticle physics both gamma (rare) and cosmic-ray (bulk) events for the training set are created by Monte Carlo simulations with approximately equal resource and time consuming. But in the real fluxes from sources in the Universe strongly prevail charged cosmic rays. In this paper, we propose to adapt the NF-based anomaly detection method to rare gamma event selection. The suggested method was tested by using the IACT data of the  TAIGA project \cite{9} and as a result we conclude that it  has a good potential for the gamma detection. However its specific implementation requires further improvements (see Section~\ref{sec:Con}).

\section{Basics on the normalizing flows\label{sec:BNF}}

A normalizing flow (NF) transforms a simple distribution (as matter of rule, the normal one) into a complex one (and vice versa) by applying a sequence (chain) of $M$ relatively simple invertible transformation functions $f_i$ and according the change of variables theorem so that eventually one obtains a probability distribution of the final target: 
\begin{eqnarray*}
p_i(\mathbf{z}_i) &=& p_{i-1}(f_i^{-1}(\mathbf{z}_i)) \left\vert \det\frac{d f_i^{-1}}{d \mathbf{z}_i} \right\vert \nonumber \\
                  &=& p_{i-1}(\mathbf{z}_{i-1}) {\left\vert \det \frac{d f_i}{d\mathbf{z}_{i-1}} \right\vert^{-1}}\qquad i=1,\dots,M\,.   
\label{eq:What_is_NF-3}
\end{eqnarray*}

Required by the computation, the transformation function $f_i$	should be: (1) easily invertible; (2) easily differentiable; (3) with the easily computable Jacobian determinant. There exist a number of models satisfying this requirements (e.g., NICE, MAF, IAF, etc. \cite{7}), most used being RealNVP \cite{10, 7}. In the latter case each $f_i$ has the form of the so called affine coupling layer. It performs an invertible transformation on the input data by splitting it into two parts, applying an affine transformation  (scaling and shifting) to one part based on the other, and leaving the other part unchanged, so that $f:\,\mathbf{x} \mapsto \mathbf{y}$ reads as follows:
\begin{eqnarray*}
\mathbf{y}_{1:d} &=& \mathbf{x}_{1:d}\ , \\ 
\mathbf{y}_{d+1:D} &=& \mathbf{x}_{d+1:D} \odot \exp({\sigma(\mathbf{x}_{1:d})}) + \mu(\mathbf{x}_{1:d})\ ,
\end{eqnarray*}
where $\sigma(\cdot)$ and $\mu(\cdot)$ are scale and translation functions and both map $\mathbb{R}^d \mapsto \mathbb{R}^{D-d}$. Here $D$ is the dimension of the data space; $d<D$. The operation $\odot$ is the element-wise product. Such a transformation is easily invertible (does not require computing the inverse of $\sigma$ or $\mu$):
\begin{eqnarray*}
\mathbf{x}_{1:d} &=& \mathbf{y}_{1:d}\ , \\ 
\mathbf{x}_{d+1:D} &=& (\mathbf{y}_{d+1:D} - \mu(\mathbf{y}_{1:d})) \odot \exp(-\sigma(\mathbf{y}_{1:d}))\ ,
\end{eqnarray*}
and thanks to the splitting the Jacobian becomes triangular and its determinant is easy to compute
$$
\mathbf{J} = 
\begin{bmatrix}
	\mathbb{I}_d & \mathbf{0}_{d\times(D-d)} \\[5pt]
	\frac{\partial \mathbf{y}_{d+1:D}}{\partial \mathbf{x}_{1:d}} & \text{diag}(\exp(\sigma(\mathbf{x}_{1:d})))
\end{bmatrix}
\qquad\Rightarrow\qquad 
\det(\mathbf{J}) = \exp(\sum_{j=1}^{D-d} \sigma(\mathbf{x}_{1:d})_j)\ .
$$
It is seen that the calculations does not involve computing the Jacobian of $\sigma$ or $\mu$, which can be arbitrarily complex. Therefore both $\sigma$ and $\mu$ can be modeled by deep neural network. During the traning stage the loss function used is the log-likelihood. 

As we indicated in the Introduction, among other applications, NFs with the RealNVP model were successfully used for anomaly detection \cite{8,11,12}. therein. In this work we suggest to use this approach for the gamma-hadron separation considering one of the particle classes as anomalies.

\section{Method and results} 
\label{sec:Method}

Briefly, the general algorithm for detecting anomalies by using the normalizing flows is as follows:
\begin{itemize}
	\item selection/extraction of essential features (dimensionality reduction) of the input data;
	\item transforming the distribution of non-anomalous data in the training set into a normal one in the latent layer by using the NF;
	\item since the mapping is optimized by means of the log-likelihood for sampling of the non-anomalous data, anomalies have a low probability in the latent space;
	\item in the input space with a complex distribution (in addition, we have only a sample) it is difficult to draw a boundary between normal and anomaly data;
	\item for a normal distribution this is easy to do, for example, the boundary can be defined via the two-sigma rule.
\end{itemize}

\begin{figure*}
	\includegraphics[scale=0.28]{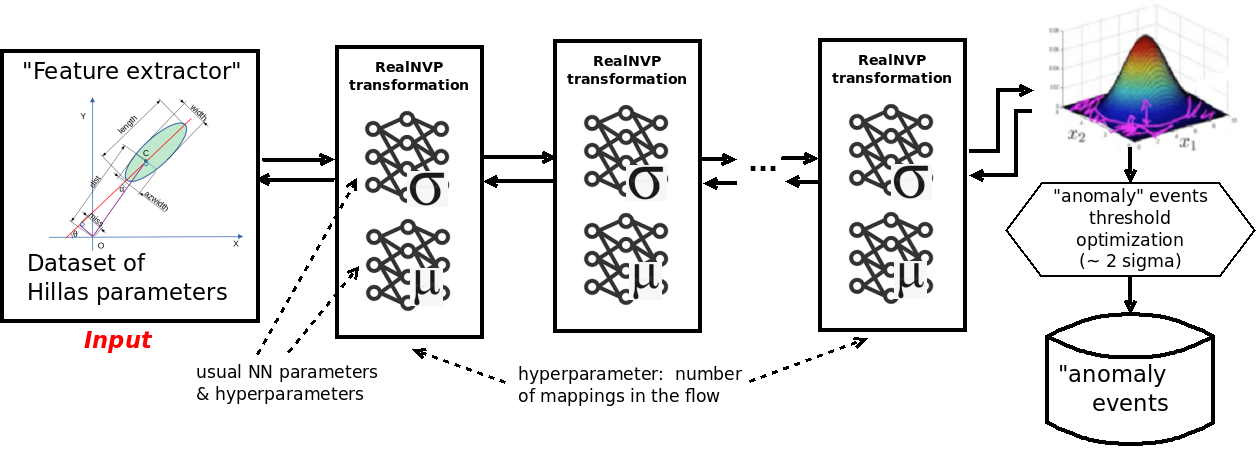}
	\caption{\label{fig:NF_scheme_anomaly} A general scheme of the gamma-hadron separation based on NF models.}
\end{figure*}

The conventional ML methods of the gamma-hadron classification are based on Monte Carlo (MC) simulated gamma-ray and hadron IACT images and two-class training \cite{2}. However, for the purposes of astrophysical research there is a need to improve IACT data processing, in particular, the particle type classification (background rejection). Therefore, its is crucial to explore all possible options to enhance the separation quality of gamma and charged particle rays. To this aim we propose to consider a chosen type of the astroparticles as anomalous events and use the NF to identify them with the two following variants: 
\begin{itemize}
	\item training set: MC hadron images; testing set: experimental $\gamma$ and hadrons images; output: low probability $\gamma$ events (considering as ''anomalies'');
	\item training set: MC $\gamma$ images; testing set: experimental $\gamma$ and hadrons images; output: low probability hadron events (''anomalies'').
\end{itemize}
As a feature extractor for the IACT data (camera images) we use the conventional Hillas algorithm \cite{2}. 
A general scheme of the gamma-hadron separation based on NF models is depicted in Fig.~\ref{fig:NF_scheme_anomaly}.
The Hillas parameters included in NF training are the following: (1) Size > 100; (2) 0.02 < Width < 0.8; (3) $\alpha$-angle (Alpha) < 20; (4) Dist < 3; (5) Length; (6) Con2. Training and validation sets contain 40000 and 500 samples, respectively. The $\mu$ and $\sigma$ functions in each of the eight NF blocks (Fig.~\ref{fig:NF_scheme_anomaly}) contain one hidden fully connected layers of 512 neurons with the ReLU activation function.  

\begin{figure*}
	\includegraphics[scale=0.32]{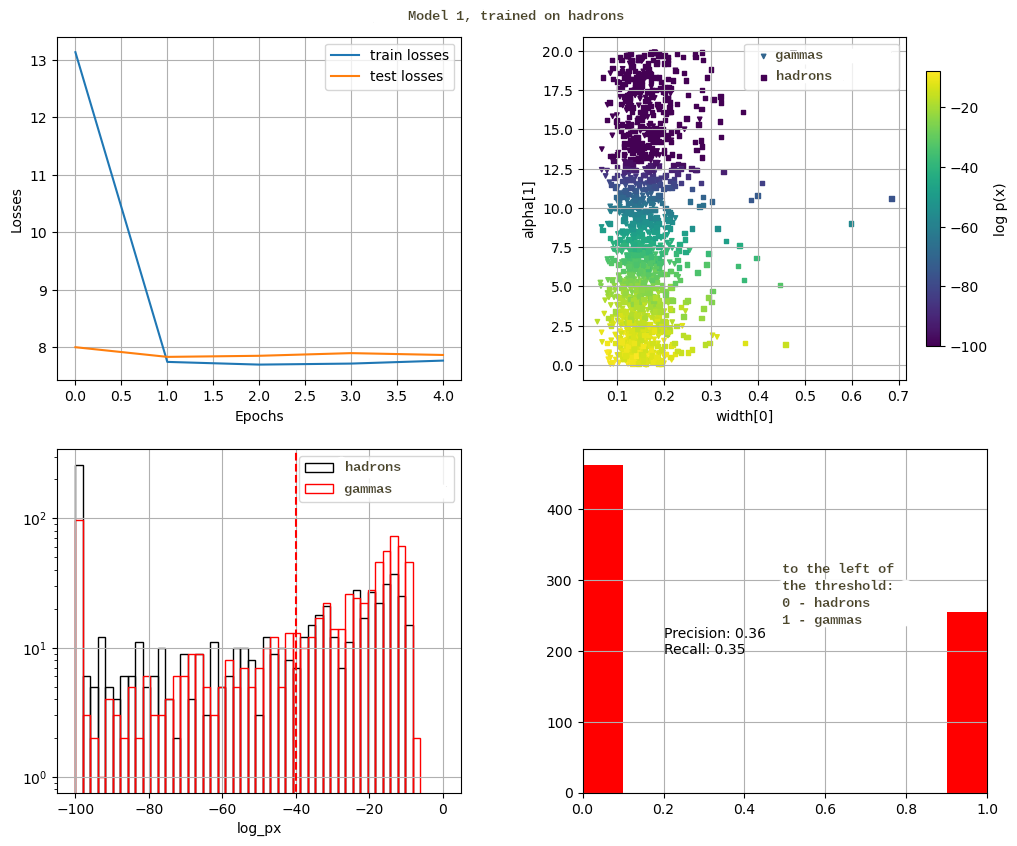}
	\caption{\label{fig:hadrons_normal} The results for the $\gamma$-event selection considering $\gamma$ as the anomalies. Top left: the loss function vs. epochs of the training. Top right: the distribution of the samples in the plane of the two Hillas parameters ($\alpha$-angle and width); the colors correspond to the probabilities of the samples in the latent space. Bottom left: the histogram of the gamma-hadron distribution by the magnitude of the log-likelihood; the chosen threshold for class separation is shown. Bottom right: distribution of data instances by classes and values of classification metrics.}
\end{figure*}

\begin{figure*}
	\includegraphics[scale=0.32]{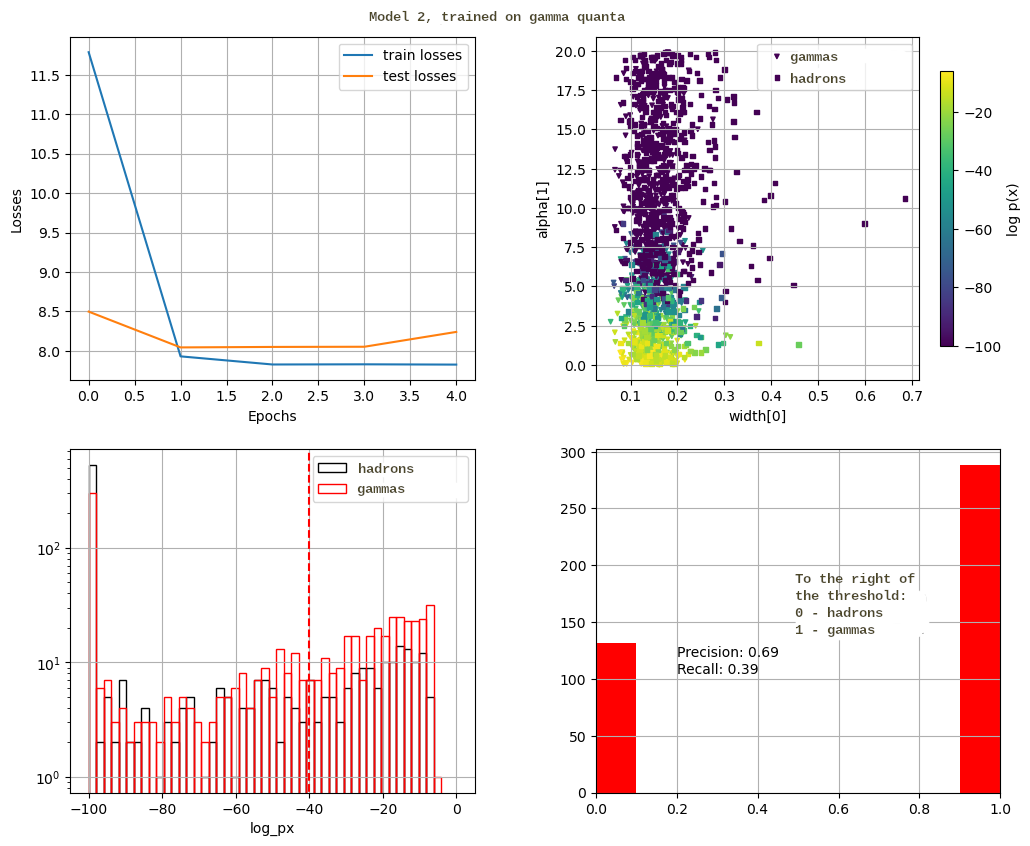}
	\caption{\label{fig:gammas_normal} The results by considering hadrons as the anomalies and the $\gamma$-events as the normal ones. Top right: the distribution of the samples in the plane of the two Hillas parameters ($\alpha$-angle and width); the colors correspond to the probabilities of the samples in the latent space. Bottom left: the histogram of the gamma-hadron distribution by the magnitude of the log-likelihood; the chosen threshold for class separation is shown. Bottom right: distribution of data instances by classes and values of classification metrics.}
\end{figure*}

\begin{figure*}
	\includegraphics[scale=0.26]{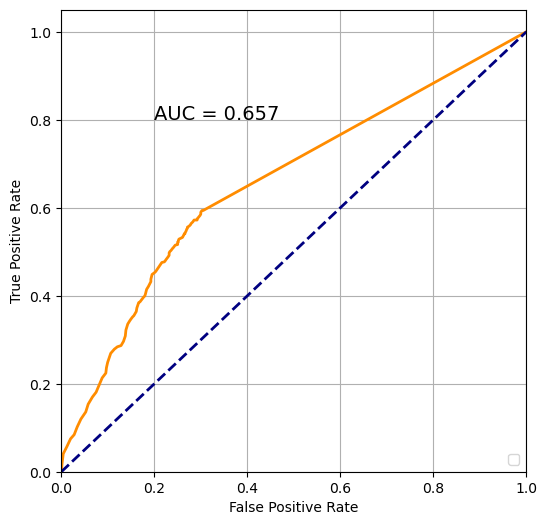}
	\caption{\label{fig:roc-auc} ROC curve for the case of hadrons as the anomalies; AUC=0.657.}
\end{figure*}

The current results for the two options are presented in Figs.~\ref{fig:hadrons_normal} and \ref{fig:gammas_normal}, respectively. ROC curve and AUC for the case of hadrons as the anomalies are presented in Fig.~\ref{fig:roc-auc}.

It is seen that the most promising use of NF is the case when hadrons, not gammas, are considered as anomalous events (Precision=0.69, Recall=0.39 in the case of  ''anomalous'' hadrons vs. Precision=0.36, Recall=0.35 for the ''anomalous'' gammas). However for the practical applications the metric values so far inappropriate for both cases and we need to improve the size and architecture of the neural networks used and perhaps the algorithm for the gamma-hadron separation on the NF basis (see the discussion in the next section).

\section{Discussion and conclusion} 
\label{sec:Con}

In conclusion, this work has been demonstrated on small neural network models that normalizing flows have the potential to isolate rare gamma events. The most promising use of NF is the case when hadrons, not gammas, are considered as anomalous events. However, the reliability of gamma isolation  obtained so far in our work is insufficient for practical use in gamma astronomy. Perhaps the reason is the small size of the training sample. Also it is necessary to use advanced neural network NF models with a larger number of transformation blocks, layers, and neurons in layers. 

It is planned to continue studying this method on more complex models and significantly larger training samples. In particular, there exists a two-class approach to the anomaly detection \cite{12}. It is based on creating synthetic samples of anomalies and training by maximizing the likelihood of normal images and minimizing the likelihood of anomalous ones. We plan to explore the possibility of applying this approach in the future. In addition, there exists a tempting possibility of reasonable using the one-class learning method. The point is that registering the EAS images with IACT out of the direction to a gamma-ray sorce, one can create with confidence a training set of purely hadron images. Using such a set of experimental hadron images as a training set, one can attempt to select gamma-ray events as anomalies in the test set. In this case, resource- and time-consuming MC simulations are not needed at all.

\begin{acknowledgments}
The work was carried out using equipment provided by Moscow University within the framework of the Moscow State University Development Program and data obtained at the unique TAIGA experimental facility.
\end{acknowledgments}

\section*{FUNDING}
This work was funded by the Russian Science Foundation (RSF), grant No. 24-11-00136, https://rscf.ru/project/24-11-00136/ .

\section*{CONFLICT OF INTEREST}
The authors of this work declare that they have no conflicts of interest.


\end{document}